\documentclass[12pt]{article}
\usepackage{graphicx}
\RequirePackage[colorlinks,citecolor=blue,urlcolor=blue,linkcolor=blue]{hyperref}
\usepackage{amsmath}
\usepackage{amssymb}
\usepackage{bold-extra}

\newcommand\pubnumber{ZU TH 03/18}
\newcommand\pubdate{\today}

\def\institute{
Physics Institute, Universit\"at Z\"urich, Z\"urich, Switzerland}
\def\support{\footnote{Work supported by the Swiss National Science Foundation~(SNF) under
contracts BSCGI0-157722 and CRSII2-160814.}}

\def\Title#1{\begin{center} {\Large #1 } \end{center}}
\def\Author#1{\begin{center}{ \sc #1} \end{center}}
\def\Address#1{\begin{center}{ \it #1} \end{center}}

\newcommand\pubblock{\rightline{\begin{tabular}{l} \pubnumber\\
         \pubdate  \end{tabular}}}
\newenvironment{Abstract}{\begin{quotation}  }{\end{quotation}}
\newenvironment{Presented}{\begin{quotation} \begin{center} 
             PRESENTED AT\end{center}\bigskip 
      \begin{center}\begin{large}}{\end{large}\end{center} \end{quotation}}
\def\Acknowledgements{\bigskip  \bigskip \begin{center} \begin{large}
             \bf ACKNOWLEDGEMENTS \end{large}\end{center}}

\def\beq{\begin{equation}}
\def\eeq#1{\label{#1}\end{equation}}
\def\eeqn{\end{equation}}

\def\beqa{\begin{eqnarray}}
\def\eeqa#1{\label{#1}\end{eqnarray}}
\def\eeqan{\end{eqnarray}}

\def\st{\scriptstyle}
\def\sst{\scriptscriptstyle}

\let\bar=\overbar

\def\Dslash{\not{\hbox{\kern-4pt $D$}}}
\def\dslash{\not{\hbox{\kern-2pt $\del$}}}

\def\mt{m_t}

\def\alphas{\alpha_s}
\def\msb{{\bar{\ssstyle M \kern -1pt S}}}

\newcommand\fourl{\ell^+\nu_{\sss\ell}\, l^-\bar{\nu}_{\sss l}}
\newcommand\bbbar{\ensuremath{b \,\bar b}}
\newcommand\sss{\mathchoice%
{\displaystyle}%
{\st}%
{\sst}%
{\sst}%
}
\newcommand{\ttbar}{\ensuremath{t \bar t}}

\newcommand\MCatNLO{MC@NLO}
\newcommand\POWHEG{\textsc{Powheg}}
\newcommand\hvq{{\tt hvq}}

\newcommand{\WWbb}{\ensuremath{W^+W^- b \bar b}}

\newcommand\POWHEGBOX{{\tt POWHEG\,BOX}}
\newcommand\bbfourl{{\tt bb4l}}
\newcommand\Pythia{{\tt Pythia}}
\newcommand\STDS{{\tt STwtDS}}
\newcommand\STDR{{\tt STwtDR}}

\newcommand{\OpenLoops}{{\tt OpenLoops}}

\newcommand{\bj}{\ensuremath{{j_{\rm \sss B}}}}

\newcommand{\kt}{\ensuremath{k_{\sss T}}}
\newcommand{\pt}{\ensuremath{p_{\sss T}}}
\newcommand{\GeV}{\mathrm{GeV}}
\newcommand{\mathd}{\mathrm{d}}
\newcommand{\tmop}[1]{\ensuremath{\operatorname{#1}}}
\def\refeq#1{\mbox{Eq.~(\ref{#1})}}
\def\refeqs#1{\mbox{Eqs.~(\ref{#1})}}
\def\reffi#1{\mbox{Fig.~\ref{#1}}}
\newcommand\pT{\ensuremath{p_{\rm T}}}
\newcommand{\nj}{\ensuremath{n_{j}}}
\newcommand{\nb}{\ensuremath{n_{b}}}
\newcommand\pTthr{\ensuremath{p_{\rm \sss T,jet}^{\rm\sss thr}}}

\begin{document}
\begin{titlepage}
\pubblock

\vfill
\Title{Top quark modelling in \POWHEGBOX{}}
\vfill
\Author{Tom\'{a}\v{s} Je\v{z}o\support}
\Address{\institute}
\vfill
\begin{Abstract}
We review recent theoretical improvements of Monte Carlo event generators for top-quark pair production and decay at the LHC based on the POWHEG method.
We present an event generator that implements spin correlations and off-shell effects in top-decay chains described in terms of exact matrix elements for $pp\to \fourl \,\bbbar$ at order $\alpha^4 \alphas^2$, including full NLO QCD corrections and interference effects with single-top and non-resonant topologies yielding to the same final state.
We then compare its predictions to previous generators that implement NLO corrections only in the top-pair production dynamics.
We consider the mass distributions of the $W\bj$ and $\ell\bj$ systems, proxies for direct top-mass determinations, and jet-vetoed cross section, a probe of the $Wt$ single top contribution.

\end{Abstract}
\vfill
\begin{Presented}
$10^{th}$ International Workshop on Top Quark Physics\\
Braga, Portugal,  September 17--22, 2017
\end{Presented}
\vfill
\end{titlepage}
\def\thefootnote{\fnsymbol{footnote}}
\setcounter{footnote}{0}

\section{Introduction}

The state-of-the art accuracy of top-pair event generators is NLO QCD, and generators matching NLO QCD matrix elements to parton showers~(NLO+PS from now on), based either upon the \MCatNLO\cite{Frixione:2002ik} or upon the \POWHEG\cite{Nason:2004rx, Frixione:2007vw} methods,  have been available for quite some time.
The first \POWHEG{} based top-pair generator\cite{Frixione:2007nw}, in the following referred to as \hvq{}, makes use of the NWA and applies NLO QCD corrections to $\ttbar$ production but not to the top-quark decay.
A generator providing NLO corrections to both production and decay, still within the NWA, was introduced in Ref.~\cite{Campbell:2014kua}.
Both generators include finite-width effects in an approximate way, the first employing the method of Ref.~\cite{Frixione:2007zp}, the second by reweighting using exact LO $pp\to \WWbb$ matrix elements including the $W$-boson decays.
A complete description of $\ttbar$ production and decay beyond the NWA requires a calculation of the full set of Feynman diagrams that contribute to the production of $\WWbb$ final states, also including leptonic or hadronic $W$-boson decays. 
%
%
A generator based on $pp\to \fourl\,\bbbar$ matrix elements, dominated by $\WWbb$ with leptonically decaying $W$-bosons, at NLO and in the 5 flavour scheme was first presented in Ref.~\cite{Garzelli:2014dka}.
However, the matching of parton showers to matrix elements that involve top-quark resonances poses nontrivial technical and theoretical problems~\cite{Jezo:2015aia} that have not been addressed in Ref.~\cite{Garzelli:2014dka}. 
A generalization of \POWHEG{} that allows for a consistent treatment of radiation from decaying resonances has been discussed in Ref.~\cite{Jezo:2015aia} and used to build a NLO+PS event generator for $pp\to \fourl\,\bbbar$ in Ref.~\cite{Jezo:2016ujg}, here referred to as \bbfourl{}.

Ref.~\cite{Jezo:2016ujg} offers a comprehensive comparison of three \POWHEGBOX{} generators with increasing accuracy of the top-decay description for both top-pair dominated and single-top enriched observables, revealing the necessity of including radiative corrections in top-quark decays.
In these proceedings, we briefly extend upon this comparison improving the predictions of \hvq{} by adding to them the contributions of single-top $tW$ topologies calculated using the generator of Ref.~\cite{Re:2010bp}.
These results are new and have not been included in other publications.
We begin with a brief review of the resonance-aware \POWHEG{} method.
We then describe the event generators used here in more detail. 
Finally, we present predictions for the mass spectra of the $W\bj$ and $\ell\bj$ systems and jet-vetoed cross sections.

\section{Resonance-aware \POWHEG{} method}

In the following we recapitulate the problems that arise in processes where intermediate narrow resonances can radiate as they decay, and summarize the ideas and methodology behind the resonance-aware algorithm of Ref.~\cite{Jezo:2015aia}. 
We refer the reader to the original publication for the description of the method in full detail.

In the \POWHEG{} method, radiation is generated according to the formula
\beq
  \mathd \sigma  =  \bar{B} (\Phi_{\mathrm{B}}) \,\mathd \Phi_{\mathrm{B}}  \Bigg[
  \Delta (q_{\tmop{cut}}) 
   +  \sum_{\alpha} \Delta (p^{\alpha}_{\sss T}) 
  \frac{R_{\alpha} (\Phi_{\alpha} (\Phi_{\mathrm{B}}, \Phi_{\tmop{rad}}))}{B
  (\Phi_{\mathrm{B}})} \,\mathd \Phi_{\tmop{rad}} \Bigg],
\eeq{eq:powheg}
where the full real matrix element has been decomposed into a sum of terms $R_{\alpha}$, each one singular only in the collinear singular region labelled by $\alpha$.
The Sudakov form factor, $\Delta$, is such that the square bracket, after performing the integrals in $\mathd \Phi_{\tmop{rad}}$, becomes exactly equal to one.
In general we have
\beq
  \Delta (q) = \prod_{\alpha} \Delta_{\alpha} (q)\,,\qquad
  \Delta_{\alpha} (q) = \exp \left[ - \int_{p^{\alpha}_{\sss T} > q}
  \frac{R_{\alpha} (\Phi_{\alpha} (\Phi_{\mathrm{B}}, \Phi_{\tmop{rad}}))}{B
  (\Phi_{\mathrm{B}})}\, \mathd \Phi_{\tmop{rad}} \right].
\eeq{eq:sudakov}
In order to achieve NLO accuracy, the $\bar{B} (\Phi_{\mathrm{B}})$ factor must equal the NLO inclusive cross section at given underlying Born kinematics.

Given the kinematics of the real-emission process and a particular singular region $\alpha$, there is a well-defined mapping that constructs a Born-like kinematic configuration $\Phi_{\mathrm{B}}$ as a function of the real one $\Phi_{\alpha} (\Phi_{\mathrm{B}}, \Phi_{\tmop{rad}})$.
This mapping is ignorant of the resonance structure of the Born process and as such does not necessarily preserve the virtuality of possible intermediate $s$-channel resonances.
This can lead to a scenario in which $R_{\alpha} / B$ terms in~\refeqs{eq:powheg} and~(\ref{eq:sudakov}) become very large, the resonances being on-shell in the numerator and off-shell in the denominator.
However, in the \POWHEG{} framework, these ratios should be either small~(of order $\alpha_s$) or should approach the Altarelli-Parisi splitting functions for the method to work.
The presence of resonances thus requires a revision of some main aspects of the \POWHEG{} method. 
Obviously, the phase space mapping $\Phi_{\alpha} (\Phi_{\mathrm{B}}, \Phi_{\tmop{rad}})$ and its inverse should preserve the virtuality of the intermediate resonances.
A modification of the singular regions decomposition is also required.
In particular, each $\alpha$ should become associated to a specific resonance structure of the event, such that collinear partons originate from the same resonance.

Such revision offers an opportunity to further improve the \POWHEG{} radiation formula.
Consider a process in which only one singular region is associated with radiation from each decaying resonance as well as with the radiation from production (i.e. not originating from a resonance decay).
This is also the case of the $g g \rightarrow (t \rightarrow W^+ b) (\bar{t} \rightarrow W^- \bar{b})$ resonance structure of the $pp\to \fourl\,\bbbar$ process.
The \POWHEG{} radiation formula can then be rewritten as:
\beq
  \mathd \sigma  =  \bar{B} (\Phi_{\mathrm{B}}) \,\mathd \Phi_{\mathrm{B}} 
  \prod_{\alpha = \alpha_{\rm\tiny prod.}, \alpha_b, \alpha_{\bar{b}}} \Bigg[
  \Delta_{\alpha} (q_{\tmop{cut}})  
 +   \Delta_{\alpha} (p^{\alpha}_{\sss T}) 
  \frac{R_{\alpha} (\Phi_{\alpha} (\Phi_{\mathrm{B}},
  \Phi^{\alpha}_{\tmop{rad}}))}{B (\Phi_{\mathrm{B}})} \,\mathd
  \Phi^{\alpha}_{\tmop{rad}} \Bigg], 
\eeq{eq:allrad}
where $\alpha_{\rm\tiny prod.}$, $\alpha_b$ and $\alpha_{\bar{b}}$ label the production singular region and the ones associated with the two top decays respectively.
Expanding the product yields a term with no emissions plus terms with multiple~(up to three) emissions. 

\section{Description of the generators}
\label{sec:pythiaInterface}

In this manuscript we consider the process of top-pair production followed by top decays in leptonic channels of different lepton families.
This corresponds to the $\fourl\,\bbbar$ final state, and we use $t\bar{t}$ and $tW$ event generators available in \POWHEGBOX{} to simulate it.
We describe them here in turn.

The first generator we consider is the most frequently used top-pair generator in the experimental analyses, the \hvq{} generator of Ref.~\cite{Frixione:2007nw}. 
It uses on-shell matrix elements for NLO production of $t\bar{t}$ pairs. 
Off-shell effects and top decays, including spin correlations, are introduced in an approximate way, according to the method presented in Ref.~\cite{Frixione:2007zp}.  
Radiation in decays is fully handled by a parton shower. 

We also consider the \bbfourl{} generator of Ref.~\cite{Jezo:2016ujg}, which implements $pp\to \fourl\,\bbbar$ matrix elements obtained with \OpenLoops{}~\cite{Cascioli:2011va}, including all QCD NLO corrections in the 4-flavour scheme, i.e.~accounting for finite $b$-mass effects.
These matrix elements are dominated by \WWbb{} topologies with leptonically decaying $W$-bosons, but also include single-top $tW$ and non-resonant %
topologies with full spin-correlation effects, radiation in production and decays, and their interference.
Thus, in comparison to the \bbfourl{} generator, the \hvq{} generator lacks contributions from non-\ttbar{} topologies. 

In order to bring \hvq{} predictions closer to those of \bbfourl{} and in order to estimate the relative impact of $tW$ topologies on the difference between the two generators we supplement \hvq{} predictions with the $tW$ contribution simulated using the generators \STDS{} and \STDR{} of Ref.~\cite{Re:2010bp}.
Similarly to \hvq{}, the \STDS{} and \STDR{} generators employ the NWA and include NLO QCD corrections only in production dynamics while they decay the top and the $W$ using the method of Ref.~\cite{Frixione:2007zp}.
The \STDS{} and \STDR{} employ the Diagram Subtraction~(DS) and Diagram Removal~(DR) schemes for removal of the \ttbar{} topologies in the real correction, respectively.

The \bbfourl{} generator can generate radiation using either the improved multiple-radiation scheme of~\refeq{eq:allrad} or the conventional single-radiation approach of~\refeq{eq:powheg}.  
In the former case the hardest radiation from all sources (i.e.~production, $t$ and $\bar{t}$ decays) may be present.
However, the standard LHIUP~\cite{Boos:2001cv} has no provisions for handling multiple instances of the hardest radiation in a single event and in order for a parton-shower to complete \bbfourl{} events consistently a non-standard interface is required.
The general idea behind such a non-standard interface is conceptually identical to the well known \pT{} veto, in other words the parton-shower algorithm is allowed to proceed without restriction, and the veto is applied if a radiation in the decaying resonance shower is harder than the \POWHEG{} generated one.
The most recent version of this interface has been described in detail in Ref.~\cite{Ravasio:2018lzi}, in which the veto is applied at the level of individual shower emissions, in contrast with the original interface of Ref.~\cite{Jezo:2016ujg}, in which the parton shower first completes the event and then reshowers it in full if vetoed.
The results in these proceedings have been obtained using the more recent interface of Ref.~\cite{Ravasio:2018lzi}.

\section{Comparison of the \bbfourl{} and the \hvq{} generators}

The calculation setup used here is identical to that of Ref.~\cite{Jezo:2016ujg} with the exception that we use the \Pythia{} interface of Ref.~\cite{Ravasio:2018lzi}, update \Pythia{} to version {\tt8.2} and enable MPI.
We consider $pp$ collisions at the center of mass energy of 8~TeV, set the top-quark mass to $\mt = 172.5$ GeV, while the value of the top-quark width is consistently calculated at NLO from all other input parameters automatically.
We denote with $B$ hadron the hardest $b$-flavoured one in the event.
Final-state hadrons are recombined into jets using the anti-$\kt$ algorithm~\cite{Cacciari:2008gp} with $R=0.5$ of {\tt FastJet}\cite{Cacciari:2011ma}.
We denote as $\bj$ the jet that contains the hardest $B$ hadron.
In the calculation of the $m_{W\bj}$ and $m_{\ell\bj}$ distributions we apply $\pt^{j} > 30~\GeV, |\eta^{j}|<2.5\,, \pt^{l} > 20~\GeV, |\eta^{l}|<2.5\,, \pt^{\sss \mathrm{miss}} > 20~\GeV$ cuts, where $\pt^{\sss\mathrm{miss}}$ is defined as the sum of the transverse momentum of final state neutrinos, while we do not apply any cuts to jet-vetoed cross sections.
A more detailed description of the setup can be found in Ref.~\cite{Jezo:2016ujg}.

\begin{figure}[htb]
\begin{center}
  \includegraphics[width=0.45\textwidth]{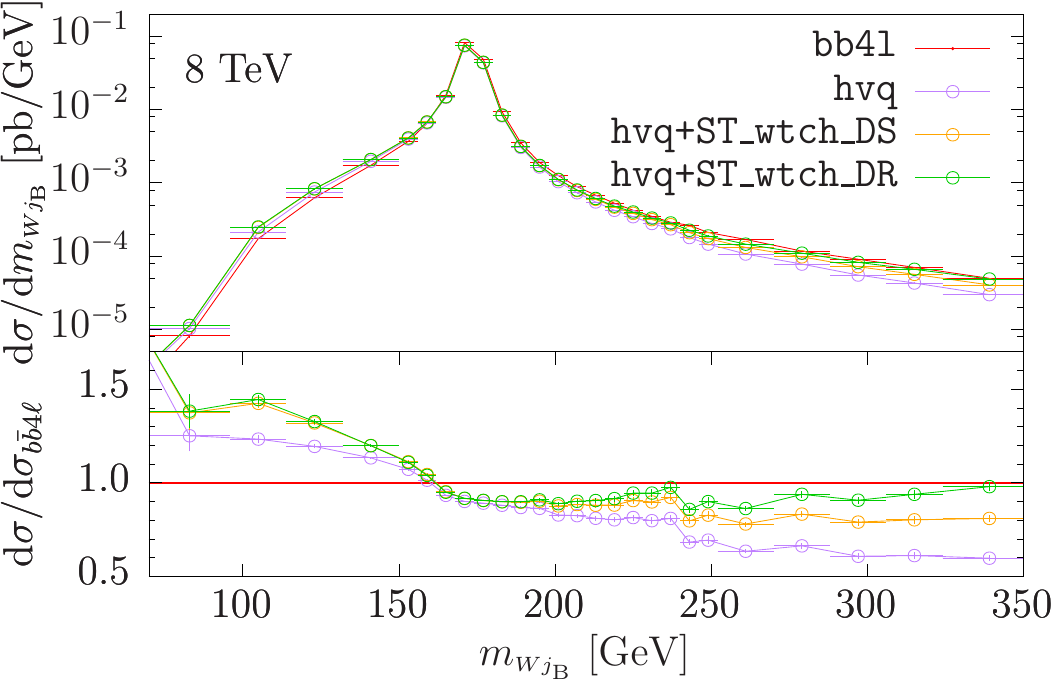}
  \includegraphics[width=0.45\textwidth]{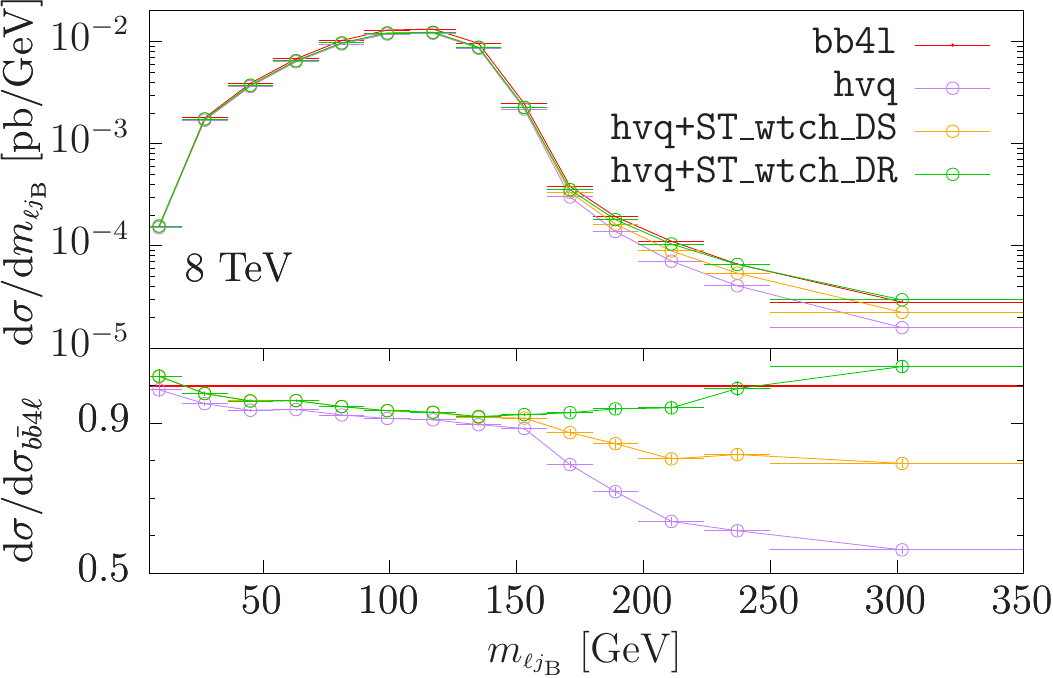}
\end{center}
\caption{
Invariant mass of the $W\bj$~(left) and of the $\ell\bj$~(right) systems.
In the ratio plot we display relative deviations with respect to the \bbfourl{} prediction.
}
\label{fig:m_w_jbot-allrad-hvq}
\end{figure}

The $W\bj$ and $\ell\bj$ mass distributions, shown in~\reffi{fig:m_w_jbot-allrad-hvq} 
show reasonably good agreement between \hvq{} and \bbfourl{} as far as the shape of the $W\bj$ peak and of the $\ell\bj$ shoulder are concerned.
However, for large top virtualities, i.e. in the tails of both distributions, sizable differences can be appreciated.
We observe that summing either \STDS{} or \STDR{} with \hvq{} leads to a considerably better agreement in the high tails of distributions.
However, the $tW$ contribution has very little impact on the shape of the $W\bj$ distribution in the vicinity of its peak.

\begin{figure}[htb]
\begin{center}
\includegraphics[width=0.45\textwidth]{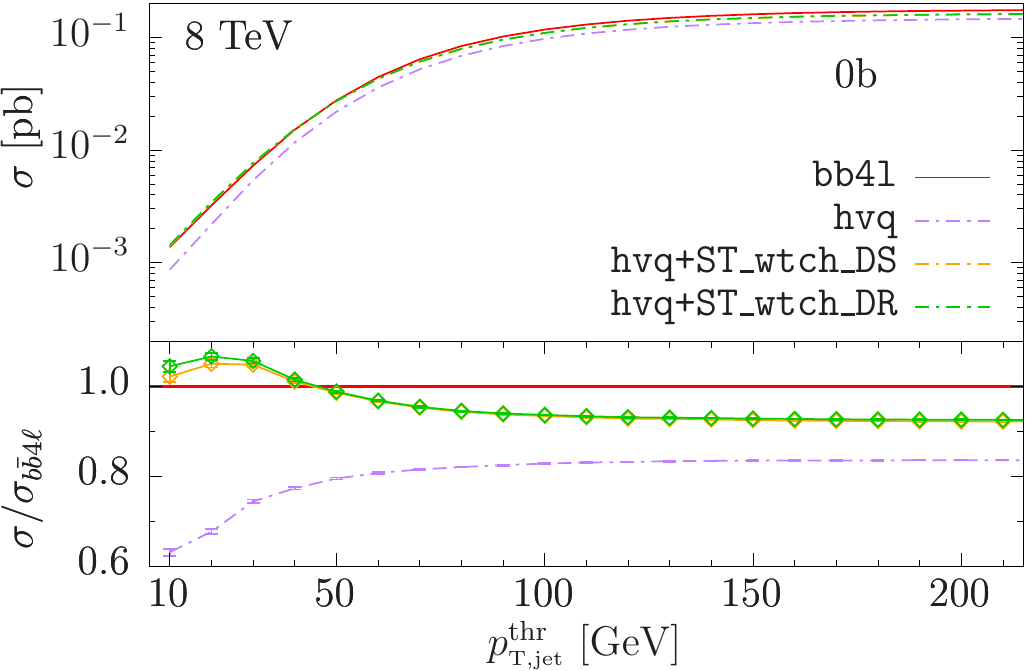}
\includegraphics[width=0.45\textwidth]{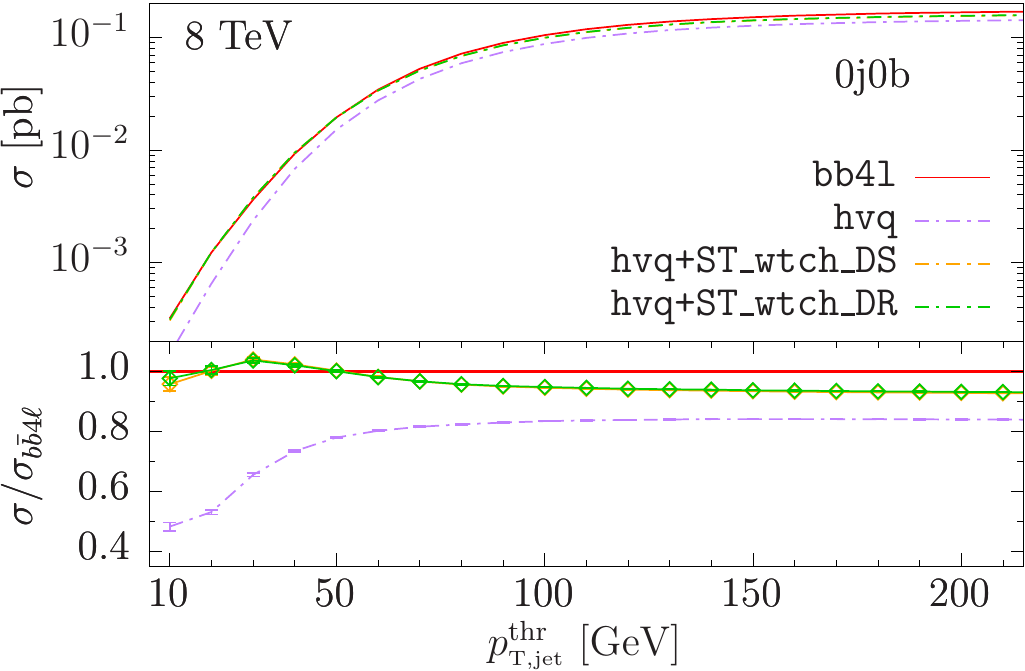}\\
\caption{
Integrated cross sections in jet bins with zero $b$-jets as a function of the jet-$\pT$ threshold.  
The left plot is inclusive with respect to extra jet radiation~($\nj{}\ge\nb{}=0$), while the right plot is exclusive ~($\nj=\nb=0$). 
The lower frame shows relative deviations with respect to the \bbfourl{} prediction. }
\label{fig:8TeV_zerob}
\end{center}
\end{figure}

Jet-binning and jet-veto effects are studied in~\reffi{fig:8TeV_zerob}.
Events are categorised according to the number of~(light \textit{or} heavy-flavour) jets, \nj{}, and to the number of $b$-jets, \nb{}, in the rapidity range $|\eta|<2.5$, while we vary the jet transverse-momentum threshold \pTthr{} that defines jets.
In the left plot the jet-veto acts only on \mbox{$b$-jets} \mbox{($\nj{}\ge\nb{}=0$),} while in the right plot a jet-veto against light and $b$-jets is applied~\mbox{($\nj{}=\nb{}=0$).}
For $\pTthr{} \gtrsim 80$~GeV the jet-vetoed cross section is dominated by \ttbar{} production and quickly converges towards the inclusive result.
In this region observe that the \hvq{} prediction features a 10\% deficit, and reducing the jet-veto scale increases this deficit up to $-30\%$ and up to $-50\%$ at $\pTthr{}=10$~GeV in the case of the inclusive $\nb{}=0$ and exclusive zero-jet cross sections~($\nj{}=\nb{}=0$, shown on the right), respectively.
However, if we add \STDS{} or \STDR{} to \hvq{}, the agreement with \bbfourl{} improves to under $10\%$ in the whole range.

\section{Summary and conclusions}
We presented the first resonance-aware NLO+PS generator for $pp\to \fourl\,\bbbar$ \cite{Jezo:2016ujg}.
We compared its predictions for $m_{W\bj}$, $m_{\ell\bj}$ and jet-vetoed cross sections to the predictions of \hvq{}, another \POWHEGBOX{} top-pair based on-shell \ttbar{} matrix elements.
We find the reliability of \hvq{} predictions decreases fast when moving away from the top resonance.
In the single-top enriched region \hvq{} fails due to missing single-top $tW$ contribution.
Supplementing \hvq{} by \STDS{} or \STDR{} improves its agreement with \bbfourl{} in high tails of $m_{W\bj}$, $m_{\ell\bj}$ distributions.
The \hvq+\STDS/\STDR{} combination does a very good job for inclusive $\nb{}=0$ and exclusive zero-jet cross sections $\nj{}=\nb{}=0$, even though it overestimates the $b$-jet \pT{} spectrum by almost $20\%$ in the low tail (not shown here).
The potential implications for the top mass determinations due to the use of the less accurate generator have been at length explored in Refs.~\cite{Ravasio:2018lzi,Nason:2018qkf}.

\Acknowledgements
I am grateful to Silvia Ferrario Ravasio and Katie Whitfield for suggestions on the manuscript.

\end{document}